\documentclass[aps,prl,superscriptaddress,reprint,longbibliography]{revtex4-1}
\usepackage{graphicx}
\usepackage{amsmath}
\usepackage{times}
\usepackage{amsfonts}
\usepackage{amssymb}
\usepackage[colorlinks=true,linkcolor=blue,citecolor=blue,urlcolor=blue]{hyperref}

\renewcommand{\Im}{\operatorname{Im}}

\newcommand{\citeasnoun}[1]{Ref.~\onlinecite{#1}}
\renewcommand{\eqref}[1]{Eq.~(\ref{eq:#1})}
\newcommand{\eqreftwo}[2]{Eqs.~(\ref{eq:#1},\ref{eq:#2})}

\newcommand{\cc}[1]{\overline{#1}}

\newcommand{\vect}[1]{\mathbf{#1}}

\begin{document}

\title{Power--bandwidth limitations of an optical resonance}

\author{Owen D. Miller}
\affiliation{Department of Mathematics, Massachusetts Institute of Technology, Cambridge, MA 02139}
\author{Chia Wei Hsu}
\affiliation{Department of Applied Physics, Yale University, New Haven, Connecticut 06520}
\author{Emma Anquillare}
\affiliation{Department of Physics, Massachusetts Institute of Technology, Cambridge, MA 02139}
\author{John D. Joannopoulos}
\affiliation{Department of Physics, Massachusetts Institute of Technology, Cambridge, MA 02139}
\author{Marin Solja\v{c}i\'{c}}
\affiliation{Department of Physics, Massachusetts Institute of Technology, Cambridge, MA 02139}
\author{Steven G. Johnson}
\affiliation{Department of Mathematics, Massachusetts Institute of Technology, Cambridge, MA 02139}

\begin{abstract}
    We present shape-independent upper limits to the power--bandwidth product for a single resonance in an optical scatterer, with the bound depending only on the material susceptibility. We show that quasistatic metallic scatterers can nearly reach the limits, and we apply our approach to the problem of designing $N$ independent, subwavelength scatterers to achieve flat, broadband response even if they individually exhibit narrow resonant peaks.
\end{abstract}

\maketitle

\date{\today}

A pivotal tool in the formulation of analytical bounds is \emph{extinction}, defined as the sum of the power absorbed and/or scattered by a body, $P_{\rm ext} = P_{\rm abs} + P_{\rm scat}$. By the optical theorem~\cite{Newton1976,Jackson1999,Lytle2005}, the extinction is given by 
\begin{align}
    P_{\rm ext} = \frac{\omega}{2} \Im \int_V \cc{\vect{E}}_{\rm inc} \cdot \vect{P},
    \label{eq:p_ext}
\end{align}
where $\vect{P}$ are the induced currents, related to the total fields through the scalar material susceptibility $\chi$, $\vect{P} = \varepsilon_0 \chi\vect{E}$, and the overline denotes complex conjugation. The integral in \eqref{p_ext} can be interpreted as a forward-scattering amplitude, since it is the overlap of the incident (forward) field with the induced currents. The fact that extinction is the imaginary part of an amplitude, instead of scaling as the squared magnitude of the fields or currents, has been the foundation of a variety of bounds~\cite{Sohl2007,Purcell1969,Pozar2009,Liberal2014,Liberal2014a,Hugonin2015,Miller2015subm,Miller2015subm2} in electromagnetism. In particular, by exploiting the fact that the integral in \eqref{p_ext} is analytic in the upper-half of the complex-frequency plane, it has been shown that the total extinction per volume, $\sigma_{\rm ext}/V$, integrated over all frequencies, is given by the electrostatic polarizability of the scatterer~\cite{Sohl2007,Purcell1969}. However, the polarizability is shape- and material-dependent, and for the common case of a metal it can be made arbitrarily large (such that the bound diverges).

Here we consider a single resonance of any scatterer, and we show that for a prescribed bandwidth there are shape-independent limits to the product of the bandwidth and the per-volume power absorbed, scattered, or extinguished by the scatterer due to the resonance under consideration. The limits depend only on the material susceptibility $\chi(\omega)$, and we show that they can be reached, e.g., by subwavelength metallic ellipsoids. The primary assumptions we make is that the resonance has a well-defined quality factor
\begin{align}
    Q = \frac{\omega U}{P_{\rm ext}},
\end{align}
which requires negligible interactions with other resonances as well as the background (as occurs for Fano lineshapes~\cite{Fan2003}), and that the stored energy can be written
\begin{align}
    U = \frac{\varepsilon_0}{2} \frac{\partial \left(\varepsilon'\omega\right)}{\partial \omega} \int_{\textrm{all space}} \left|\vect{E}\right|^2
    \label{eq:u_tot}
\end{align}
where $\varepsilon'$ is the real part of the permittivity, which is a common expression~\cite{Wang2006} but is rigorously valid only in the low-loss limit~\cite{Raman2013}. We consider the total $Q$, which comprises radiative and absorptive components, $1/Q = 1/Q_{\rm abs} + 1/Q_{\rm scat}$.

We consider plane-wave scattering throughout (generalizations are straightfoward), such that a more natural quantity is the cross-section, $\sigma$, defined for any power quantity as the power $P$ divided by the incident intensity, $I_{\rm inc} = |E_0|^2 / 2Z_0$, where $E_0$ is the plane-wave amplitude and $Z_0$ is the impedance of free space. The integral of the per-volume cross-section associated with the single resonance is given by the product of the peak extinction at some wavelength $\lambda_0$, and the resonant bandwidth $\Delta\lambda$:
\begin{align}
    \frac{1}{V} \int \sigma_{\rm ext} \,{\rm d}\lambda = \frac{\pi}{2} \frac{\sigma_{\rm ext}(\lambda_0)}{V} \Delta \lambda 
    \label{eq:dl_def}
\end{align}
where \eqref{dl_def} serves as the \emph{definition} of the bandwidth $\Delta\lambda$. We have included the factor of $\pi/2$ so that in the common case of a Lorentzian extinction lineshape, our bandwidth $\Delta\lambda$ is precisely the full-width at half-maximum of the Lorentzian. We primarily use wavelength instead of frequency so that quantities as in \eqref{dl_def} are dimensionless, but the translation to frequency bounds is simple.

The limits follow from variational calculus arguments. Because $\Delta\lambda = \lambda_0 / Q$, one can rewrite \eqref{dl_def} as
\begin{align}
    \frac{1}{V} \int \sigma_{\rm ext} \,{\rm d}\lambda = \frac{\lambda_0^2}{4cI_{\rm inc}V} \frac{P_{\rm ext}^2(\lambda_0)}{U(\lambda_0)} 
    \label{eq:ext_int}
\end{align}
The total stored energy is greater than or equal to energy stored in the interior of the scatterer, $U > U_{\rm int}$, where
\begin{align}
    U_{\rm int} = \frac{\varepsilon_0}{2} \frac{\partial \left(\varepsilon'\omega\right)}{\partial \omega} \int_V \left|\vect{E}\right|^2.
    \label{eq:u_int}
\end{align}
Then the quantity $P_{\rm ext}^2/U$ is bounded above by
\begin{align}
    \frac{P_{\rm ext}^2(\lambda_0)}{U(\lambda_0)} \leq \frac{2\pi^2}{\mu_0 \lambda_0^2} \left|\chi\right|^2 \left[ \frac{\partial\left(\varepsilon'\omega\right)}{\partial \omega} \right]^{-1} \frac{\left[ \Im \int_V \cc{\vect{E}_{\rm inc}} \cdot \vect{P} \right]^2}{\int_V \left|\vect{P}\right|^2}
    \label{eq:p2_over_u}
\end{align}
where we have rewritten the (interior) energy density in terms of the polarization currents. \eqref{p2_over_u} has both a numerator and denominator that is quadratic in the polarization current density, $\vect{P}$, such that it cannot be made arbitrarily large. A quick variational derivative (taking $\partial / \partial \vect{P}(\vect{x})$, for some $\vect{x}$, in essence) shows that the optimal polarization current to maximize \eqref{p2_over_u} is given by
\begin{align}
    \vect{P}_{\rm opt} = ai\vect{E}_{\rm inc}
\end{align}
where $a$ is any real constant and this choice maximizes the overlap between the incident field and the induced currents. For the optimal currents, the quantity $\left[ \Im \int_V \cc{\vect{E}_{\rm inc}} \cdot \vect{P} \right]^2 / \int_V \left|\vect{P}\right|^2 = |E_0|^2 V$, simplifying the inequality in \eqref{p2_over_u} and leading to the bound on the integrated per-volume extinction:
\begin{align}
    \frac{1}{V} \int \sigma_{\rm ext} \,{\rm d}\lambda \leq \pi^2 \left. \frac{\left|\chi\right|^2}{\frac{\partial\left(\varepsilon'\omega\right)}{\partial \omega}} \right|_{\lambda_0}
    \label{eq:ext_bound}
\end{align}
For a material with negligible dispersion over the bandwidth of interest, and a moderate to large susceptibility (e.g. $\chi=9$, as for a material with refractive index $n=3$) \eqref{ext_bound} becomes 
\begin{align}
    \frac{1}{V} \int \sigma_{\rm ext} \,{\rm d}\lambda &\leq \pi^2 \left. \frac{\left|\chi\right|^2}{\varepsilon'} \right|_{\lambda_0} \nonumber \\
                                                       &\lesssim \pi^2 \left|\chi\right| \label{eq:ext_bound_diel}
\end{align}
\eqreftwo{ext_bound}{ext_bound_diel} form the central results of this Letter. They show that the broadband response associated with a given resonance in a scatterer is bounded by a simple function of the material susceptibility $\chi$ at the resonant wavelength $\lambda_0$.

\begin{figure}
\centering
\includegraphics[width=\linewidth]{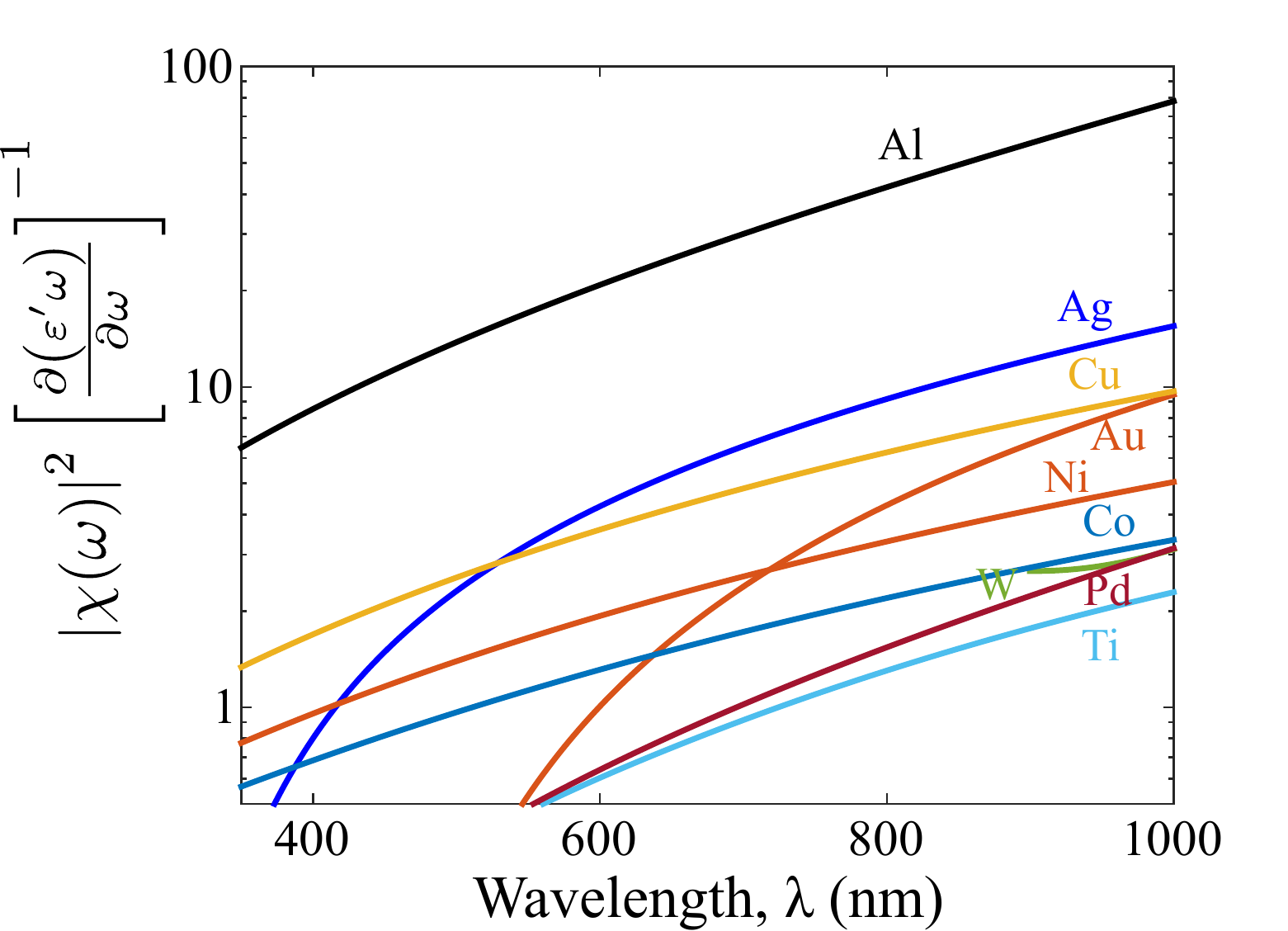}
\caption{\label{fig:fig1}Comparison of the material parameter that determines the bound on the power--bandwidth product of a single resonance in any material. The material susceptibilities are fit to experimental data with Lorentz-Drude oscillator terms. Aluminum is ideal for broadband applications due to its large susceptibility.}
\end{figure}

The limits are nearly attainable with quasistatic metallic ellipsoids. For a Drude metal with $\chi = -\omega_p^2 / (\omega^2 + i\gamma\omega)$, for plasma frequency $\omega_p$ and damping factor $\gamma$, the bound in \eqref{ext_bound} is given by
\begin{align}
    \frac{1}{V} \int \sigma_{\rm ext} \,{\rm d}\lambda \leq \pi^2 \frac{\omega_p^4 \left(\omega^2+\gamma^2\right)}{\omega^2\left[\left(\omega^2+\gamma^2\right)^2 +\omega^2\omega_p^2 - \omega_p^2\gamma^2\right]}
    \label{eq:ext_bound_drude}
\end{align}
A quasistatic ellipsoid has a peak extinction cross-section per volume of $\omega_p^2 / \gamma c$ and a quality factor $Q = \omega/\gamma$ (\citeasnoun{Raman2013}) in a Drude model~\cite{Bohren1983}, such that the power--bandwidth integral is given by
\begin{align}
    \left[\frac{1}{V} \int \sigma_{\rm ext} \,{\rm d}\lambda\right]_{\textrm{ellipsoid,qs}} = \pi^2 \frac{\omega_p^2}{\omega^2}.
    \label{eq:ext_ell_qs}
\end{align}
In the low-loss, low-frequency regime, $\gamma \ll \omega \ll \omega_p$, \eqref{ext_bound_drude} approximately reduces to \eqref{ext_ell_qs}.

For a collection of $N$ single-mode scatterers (e.g. subwavelength particles~\cite{Anquillare2015subm}) that absorb and scatter light independently, the results here can be used to define the optimal \emph{maximin} cross-section per volume, i.e. the maximum ``worst-case'' extinction per volume over a desired bandwidth $\lambda_2 - \lambda_1$. To construct the ideal lineshape with a flat response between $\lambda_1$ and $\lambda_2$, and zero elsewhere, one would ideally have each resonance take such a rectangular form as well (although unphysical, this enables us to find the upper limit). Then, for any combination of resonant peaks and bandwidths, the largest possible amplitude between $\lambda_1$ and $\lambda_2$ is given by the average value of \eqref{ext_bound}, leading to the maximin bound
\begin{align}
    \left[\frac{\sigma_{\rm ext}}{V}\right]_{\rm maximin} \leq \frac{\pi^2}{\lambda_2 - \lambda_1} 
    \left[ \frac{\left|\chi\right|^2}{\frac{\partial\left(\varepsilon'\omega\right)}{\partial \omega}} \right]_{\textrm{avg,}\lambda_1\textrm{--}\lambda_2}
    \label{eq:maximin_bound}
\end{align}

Intuitively, the bound presented here can be understood through the polarization currents. Because \eqref{p_ext} is written only over the volume of the scatterer, any significant absorption or scattering peak must be accompanied by large polarization within the scatterer. Such large current densities in a single mode necessarily implies a large stored energy, and thus a high quality factor and a narrow bandwidth.

\bibliography{/home/odmiller/texmf/bibtex/bib/library}

\begin{thebibliography}{16}%
\makeatletter
\providecommand \@ifxundefined [1]{%
 \@ifx{#1\undefined}
}%
\providecommand \@ifnum [1]{%
 \ifnum #1\expandafter \@firstoftwo
 \else \expandafter \@secondoftwo
 \fi
}%
\providecommand \@ifx [1]{%
 \ifx #1\expandafter \@firstoftwo
 \else \expandafter \@secondoftwo
 \fi
}%
\providecommand \natexlab [1]{#1}%
\providecommand \enquote  [1]{``#1''}%
\providecommand \bibnamefont  [1]{#1}%
\providecommand \bibfnamefont [1]{#1}%
\providecommand \citenamefont [1]{#1}%
\providecommand \href@noop [0]{\@secondoftwo}%
\providecommand \href [0]{\begingroup \@sanitize@url \@href}%
\providecommand \@href[1]{\@@startlink{#1}\@@href}%
\providecommand \@@href[1]{\endgroup#1\@@endlink}%
\providecommand \@sanitize@url [0]{\catcode `\\12\catcode `\$12\catcode
  `\&12\catcode `\#12\catcode `\^12\catcode `\_12\catcode `\%12\relax}%
\providecommand \@@startlink[1]{}%
\providecommand \@@endlink[0]{}%
\providecommand \url  [0]{\begingroup\@sanitize@url \@url }%
\providecommand \@url [1]{\endgroup\@href {#1}{\urlprefix }}%
\providecommand \urlprefix  [0]{URL }%
\providecommand \Eprint [0]{\href }%
\providecommand \doibase [0]{http://dx.doi.org/}%
\providecommand \selectlanguage [0]{\@gobble}%
\providecommand \bibinfo  [0]{\@secondoftwo}%
\providecommand \bibfield  [0]{\@secondoftwo}%
\providecommand \translation [1]{[#1]}%
\providecommand \BibitemOpen [0]{}%
\providecommand \bibitemStop [0]{}%
\providecommand \bibitemNoStop [0]{.\EOS\space}%
\providecommand \EOS [0]{\spacefactor3000\relax}%
\providecommand \BibitemShut  [1]{\csname bibitem#1\endcsname}%
\let\auto@bib@innerbib\@empty
\bibitem [{\citenamefont {Newton}(1976)}]{Newton1976}%
  \BibitemOpen
  \bibfield  {author} {\bibinfo {author} {\bibfnamefont {Roger~G.}\
  \bibnamefont {Newton}},\ }\bibfield  {title} {\enquote {\bibinfo {title}
  {{Optical theorem and beyond}},}\ }\href {\doibase 10.1119/1.10324}
  {\bibfield  {journal} {\bibinfo  {journal} {Am. J. Phys.}\ }\textbf {\bibinfo
  {volume} {44}},\ \bibinfo {pages} {639} (\bibinfo {year} {1976})}\BibitemShut
  {NoStop}%
\bibitem [{\citenamefont {Jackson}(1999)}]{Jackson1999}%
  \BibitemOpen
  \bibfield  {author} {\bibinfo {author} {\bibfnamefont {J.~D.}\ \bibnamefont
  {Jackson}},\ }\href@noop {} {\emph {\bibinfo {title} {{Classical
  Electrodynamics, 3rd Ed.}}}}\ (\bibinfo  {publisher} {John Wiley \& Sons,
  Inc.},\ \bibinfo {year} {1999})\BibitemShut {NoStop}%
\bibitem [{\citenamefont {Lytle}\ \emph {et~al.}(2005)\citenamefont {Lytle},
  \citenamefont {Carney}, \citenamefont {Schotland},\ and\ \citenamefont
  {Wolf}}]{Lytle2005}%
  \BibitemOpen
  \bibfield  {author} {\bibinfo {author} {\bibfnamefont {D.~R.}\ \bibnamefont
  {Lytle}}, \bibinfo {author} {\bibfnamefont {P.~Scott}\ \bibnamefont
  {Carney}}, \bibinfo {author} {\bibfnamefont {John~C.}\ \bibnamefont
  {Schotland}}, \ and\ \bibinfo {author} {\bibfnamefont {Emil}\ \bibnamefont
  {Wolf}},\ }\bibfield  {title} {\enquote {\bibinfo {title} {{Generalized
  optical theorem for reflection, transmission, and extinction of power for
  electromagnetic fields}},}\ }\href {\doibase 10.1103/PhysRevE.71.056610}
  {\bibfield  {journal} {\bibinfo  {journal} {Phys. Rev. E}\ }\textbf {\bibinfo
  {volume} {71}},\ \bibinfo {pages} {056610} (\bibinfo {year}
  {2005})}\BibitemShut {NoStop}%
\bibitem [{\citenamefont {Sohl}\ \emph {et~al.}(2007)\citenamefont {Sohl},
  \citenamefont {Gustafsson},\ and\ \citenamefont {Kristensson}}]{Sohl2007}%
  \BibitemOpen
  \bibfield  {author} {\bibinfo {author} {\bibfnamefont {Christian}\
  \bibnamefont {Sohl}}, \bibinfo {author} {\bibfnamefont {Mats}\ \bibnamefont
  {Gustafsson}}, \ and\ \bibinfo {author} {\bibfnamefont {Gerhard}\
  \bibnamefont {Kristensson}},\ }\bibfield  {title} {\enquote {\bibinfo {title}
  {{Physical limitations on metamaterials: restrictions on scattering and
  absorption over a frequency interval}},}\ }\href {\doibase
  10.1088/0022-3727/40/22/042} {\bibfield  {journal} {\bibinfo  {journal} {J.
  Phys. D. Appl. Phys.}\ }\textbf {\bibinfo {volume} {40}},\ \bibinfo {pages}
  {7146--7151} (\bibinfo {year} {2007})}\BibitemShut {NoStop}%
\bibitem [{\citenamefont {Purcell}(1969)}]{Purcell1969}%
  \BibitemOpen
  \bibfield  {author} {\bibinfo {author} {\bibfnamefont {E.~M.}\ \bibnamefont
  {Purcell}},\ }\bibfield  {title} {\enquote {\bibinfo {title} {{On the
  Absorption and Emission of Light by Interstellar Grains}},}\ }\href {\doibase
  10.1086/150207} {\bibfield  {journal} {\bibinfo  {journal} {Astrophys. J.}\
  }\textbf {\bibinfo {volume} {158}},\ \bibinfo {pages} {433--440} (\bibinfo
  {year} {1969})}\BibitemShut {NoStop}%
\bibitem [{\citenamefont {Kwon}\ and\ \citenamefont {Pozar}(2009)}]{Pozar2009}%
  \BibitemOpen
  \bibfield  {author} {\bibinfo {author} {\bibfnamefont {Do-Hoon}\ \bibnamefont
  {Kwon}}\ and\ \bibinfo {author} {\bibfnamefont {David~M.}\ \bibnamefont
  {Pozar}},\ }\bibfield  {title} {\enquote {\bibinfo {title} {{Optimal
  Characteristics of an Arbitrary Receive Antenna}},}\ }\href {\doibase
  10.1109/TAP.2009.2025975} {\bibfield  {journal} {\bibinfo  {journal} {IEEE
  Trans. Antennas Propag.}\ }\textbf {\bibinfo {volume} {57}},\ \bibinfo
  {pages} {3720--3727} (\bibinfo {year} {2009})}\BibitemShut {NoStop}%
\bibitem [{\citenamefont {Liberal}\ \emph
  {et~al.}(2014{\natexlab{a}})\citenamefont {Liberal}, \citenamefont {Ra'di},
  \citenamefont {Gonzalo}, \citenamefont {Ederra}, \citenamefont {Tretyakov},\
  and\ \citenamefont {Ziolkowski}}]{Liberal2014}%
  \BibitemOpen
  \bibfield  {author} {\bibinfo {author} {\bibfnamefont {Inigo}\ \bibnamefont
  {Liberal}}, \bibinfo {author} {\bibfnamefont {Younes}\ \bibnamefont {Ra'di}},
  \bibinfo {author} {\bibfnamefont {Ramon}\ \bibnamefont {Gonzalo}}, \bibinfo
  {author} {\bibfnamefont {Inigo}\ \bibnamefont {Ederra}}, \bibinfo {author}
  {\bibfnamefont {Sergei~A.}\ \bibnamefont {Tretyakov}}, \ and\ \bibinfo
  {author} {\bibfnamefont {Richard~W.}\ \bibnamefont {Ziolkowski}},\ }\bibfield
   {title} {\enquote {\bibinfo {title} {{Least Upper Bounds of the Powers
  Extracted and Scattered by Bi-anisotropic Particles}},}\ }\href {\doibase
  10.1109/TAP.2014.2330620} {\bibfield  {journal} {\bibinfo  {journal} {IEEE
  Trans. Antennas Propag.}\ }\textbf {\bibinfo {volume} {62}},\ \bibinfo
  {pages} {4726} (\bibinfo {year} {2014}{\natexlab{a}})},\ \Eprint
  {http://arxiv.org/abs/1402.4726} {arXiv:1402.4726} \BibitemShut {NoStop}%
\bibitem [{\citenamefont {Liberal}\ \emph
  {et~al.}(2014{\natexlab{b}})\citenamefont {Liberal}, \citenamefont {Ederra},
  \citenamefont {Gonzalo},\ and\ \citenamefont {Ziolkowski}}]{Liberal2014a}%
  \BibitemOpen
  \bibfield  {author} {\bibinfo {author} {\bibfnamefont {Inigo}\ \bibnamefont
  {Liberal}}, \bibinfo {author} {\bibfnamefont {Inigo}\ \bibnamefont {Ederra}},
  \bibinfo {author} {\bibfnamefont {Ramon}\ \bibnamefont {Gonzalo}}, \ and\
  \bibinfo {author} {\bibfnamefont {Richard}\ \bibnamefont {Ziolkowski}},\
  }\bibfield  {title} {\enquote {\bibinfo {title} {{Upper Bounds on Scattering
  Processes and Metamaterial-Inspired Structures that Reach Them}},}\ }\href
  {\doibase 10.1109/TAP.2014.2359206} {\bibfield  {journal} {\bibinfo
  {journal} {IEEE Trans. Antennas Propag.}\ ,\ \bibinfo {pages} {6344}}
  (\bibinfo {year} {2014}{\natexlab{b}})}\BibitemShut {NoStop}%
\bibitem [{\citenamefont {Hugonin}\ \emph {et~al.}(2015)\citenamefont
  {Hugonin}, \citenamefont {Besbes},\ and\ \citenamefont
  {Ben-Abdallah}}]{Hugonin2015}%
  \BibitemOpen
  \bibfield  {author} {\bibinfo {author} {\bibfnamefont {Jean-Paul}\
  \bibnamefont {Hugonin}}, \bibinfo {author} {\bibfnamefont {Mondher}\
  \bibnamefont {Besbes}}, \ and\ \bibinfo {author} {\bibfnamefont {Philippe}\
  \bibnamefont {Ben-Abdallah}},\ }\bibfield  {title} {\enquote {\bibinfo
  {title} {{Fundamental limits for light absorption and scattering induced by
  cooperative electromagnetic interactions}},}\ }\href {\doibase
  10.1103/PhysRevB.91.180202} {\bibfield  {journal} {\bibinfo  {journal} {Phys.
  Rev. B}\ }\textbf {\bibinfo {volume} {91}},\ \bibinfo {pages} {180202}
  (\bibinfo {year} {2015})},\ \Eprint {http://arxiv.org/abs/1502.02409}
  {arXiv:1502.02409} \BibitemShut {NoStop}%
\bibitem [{\citenamefont {Miller}\ \emph
  {et~al.}(2015{\natexlab{a}})\citenamefont {Miller}, \citenamefont
  {Polimeridis}, \citenamefont {Reid}, \citenamefont {Hsu}, \citenamefont
  {DeLacy}, \citenamefont {Joannopoulos}, \citenamefont {Solja\v{c}i\'{c}},\
  and\ \citenamefont {Johnson}}]{Miller2015subm}%
  \BibitemOpen
  \bibfield  {author} {\bibinfo {author} {\bibfnamefont {Owen~D.}\ \bibnamefont
  {Miller}}, \bibinfo {author} {\bibfnamefont {Athanasios~G.}\ \bibnamefont
  {Polimeridis}}, \bibinfo {author} {\bibfnamefont {M.~T.~Homer}\ \bibnamefont
  {Reid}}, \bibinfo {author} {\bibfnamefont {Chia~Wei}\ \bibnamefont {Hsu}},
  \bibinfo {author} {\bibfnamefont {Brendan~G.}\ \bibnamefont {DeLacy}},
  \bibinfo {author} {\bibfnamefont {John~D.}\ \bibnamefont {Joannopoulos}},
  \bibinfo {author} {\bibfnamefont {Marin}\ \bibnamefont {Solja\v{c}i\'{c}}}, \
  and\ \bibinfo {author} {\bibfnamefont {Steven~G.}\ \bibnamefont {Johnson}},\
  }\bibfield  {title} {\enquote {\bibinfo {title} {{Fundamental limits to the
  optical response of lossy media}},}\ }\href@noop {} {\bibfield  {journal}
  {\bibinfo  {journal} {Submitt. arXiv 1503.03781}\ } (\bibinfo {year}
  {2015}{\natexlab{a}})}\BibitemShut {NoStop}%
\bibitem [{\citenamefont {Miller}\ \emph
  {et~al.}(2015{\natexlab{b}})\citenamefont {Miller}, \citenamefont {Johnson},\
  and\ \citenamefont {Rodriguez}}]{Miller2015subm2}%
  \BibitemOpen
  \bibfield  {author} {\bibinfo {author} {\bibfnamefont {Owen~D.}\ \bibnamefont
  {Miller}}, \bibinfo {author} {\bibfnamefont {Steven~G.}\ \bibnamefont
  {Johnson}}, \ and\ \bibinfo {author} {\bibfnamefont {Alejandro~W.}\
  \bibnamefont {Rodriguez}},\ }\bibfield  {title} {\enquote {\bibinfo {title}
  {{Shape-independent limits to near-field radiative heat transfer}},}\ }\href
  {http://arxiv.org/abs/1504.01323} {\bibfield  {journal} {\bibinfo  {journal}
  {arXiv: 1504.01323}\ } (\bibinfo {year} {2015}{\natexlab{b}})},\ \Eprint
  {http://arxiv.org/abs/1504.01323} {arXiv:1504.01323} \BibitemShut {NoStop}%
\bibitem [{\citenamefont {Fan}\ \emph {et~al.}(2003)\citenamefont {Fan},
  \citenamefont {Suh},\ and\ \citenamefont {Joannopoulos}}]{Fan2003}%
  \BibitemOpen
  \bibfield  {author} {\bibinfo {author} {\bibfnamefont {Shanhui}\ \bibnamefont
  {Fan}}, \bibinfo {author} {\bibfnamefont {Wonjoo}\ \bibnamefont {Suh}}, \
  and\ \bibinfo {author} {\bibfnamefont {J.~D.}\ \bibnamefont {Joannopoulos}},\
  }\bibfield  {title} {\enquote {\bibinfo {title} {{Temporal coupled-mode
  theory for the Fano resonance in optical resonators}},}\ }\href {\doibase
  10.1364/JOSAA.20.000569} {\bibfield  {journal} {\bibinfo  {journal} {J. Opt.
  Soc. Am. A}\ }\textbf {\bibinfo {volume} {20}},\ \bibinfo {pages} {569}
  (\bibinfo {year} {2003})}\BibitemShut {NoStop}%
\bibitem [{\citenamefont {Wang}\ and\ \citenamefont {Wang}(2006)}]{Wang2006}%
  \BibitemOpen
  \bibfield  {author} {\bibinfo {author} {\bibfnamefont {Shengyin}\
  \bibnamefont {Wang}}\ and\ \bibinfo {author} {\bibfnamefont {Michael~Yu}\
  \bibnamefont {Wang}},\ }\bibfield  {title} {\enquote {\bibinfo {title}
  {{Radial basis functions and level set method for structural topology
  optimization}},}\ }\href {\doibase 10.1002/nme.1536} {\bibfield  {journal}
  {\bibinfo  {journal} {Int. J. Numer. Methods Eng.}\ }\textbf {\bibinfo
  {volume} {65}},\ \bibinfo {pages} {2060--2090} (\bibinfo {year}
  {2006})}\BibitemShut {NoStop}%
\bibitem [{\citenamefont {Raman}\ \emph {et~al.}(2013)\citenamefont {Raman},
  \citenamefont {Shin},\ and\ \citenamefont {Fan}}]{Raman2013}%
  \BibitemOpen
  \bibfield  {author} {\bibinfo {author} {\bibfnamefont {Aaswath}\ \bibnamefont
  {Raman}}, \bibinfo {author} {\bibfnamefont {Wonseok}\ \bibnamefont {Shin}}, \
  and\ \bibinfo {author} {\bibfnamefont {Shanhui}\ \bibnamefont {Fan}},\
  }\bibfield  {title} {\enquote {\bibinfo {title} {{Upper Bound on the Modal
  Material Loss Rate in Plasmonic and Metamaterial Systems}},}\ }\href
  {\doibase 10.1103/PhysRevLett.110.183901} {\bibfield  {journal} {\bibinfo
  {journal} {Phys. Rev. Lett.}\ }\textbf {\bibinfo {volume} {110}},\ \bibinfo
  {pages} {183901} (\bibinfo {year} {2013})}\BibitemShut {NoStop}%
\bibitem [{\citenamefont {Bohren}\ and\ \citenamefont
  {Huffman}(1983)}]{Bohren1983}%
  \BibitemOpen
  \bibfield  {author} {\bibinfo {author} {\bibfnamefont {Craig~F.}\
  \bibnamefont {Bohren}}\ and\ \bibinfo {author} {\bibfnamefont {Donald~R.}\
  \bibnamefont {Huffman}},\ }\href@noop {} {\emph {\bibinfo {title}
  {{Absorption and Scattering of Light by Small Particles}}}}\ (\bibinfo
  {publisher} {John Wiley \& Sons},\ \bibinfo {address} {New York, NY},\
  \bibinfo {year} {1983})\BibitemShut {NoStop}%
\bibitem [{\citenamefont {Anquillare}\ \emph {et~al.}(2015)\citenamefont
  {Anquillare}, \citenamefont {Miller}, \citenamefont {Hsu}, \citenamefont
  {DeLacy}, \citenamefont {Joannopoulos}, \citenamefont {Johnson},\ and\
  \citenamefont {Solja\v{c}i\'{c}}}]{Anquillare2015subm}%
  \BibitemOpen
  \bibfield  {author} {\bibinfo {author} {\bibfnamefont {Emma}\ \bibnamefont
  {Anquillare}}, \bibinfo {author} {\bibfnamefont {Owen~D.}\ \bibnamefont
  {Miller}}, \bibinfo {author} {\bibfnamefont {Chia~Wei}\ \bibnamefont {Hsu}},
  \bibinfo {author} {\bibfnamefont {Brendan~G.}\ \bibnamefont {DeLacy}},
  \bibinfo {author} {\bibfnamefont {John~D.}\ \bibnamefont {Joannopoulos}},
  \bibinfo {author} {\bibfnamefont {Steven~G.}\ \bibnamefont {Johnson}}, \ and\
  \bibinfo {author} {\bibfnamefont {Marin}\ \bibnamefont {Solja\v{c}i\'{c}}},\
  }\bibfield  {title} {\enquote {\bibinfo {title} {{Efficient broad- and
  tunable-bandwidth optical extinction via aspect-ratio-tailored silver
  nanodisks}},}\ }\href@noop {} {\bibfield  {journal} {\bibinfo  {journal}
  {Submitted}\ } (\bibinfo {year} {2015})},\ \Eprint
  {http://arxiv.org/abs/1510.01768} {arXiv:1510.01768} \BibitemShut {NoStop}%
\end{thebibliography}%
\end{document}